\documentclass[a4paper]{article}
\RequirePackage[english]{babel}
\RequirePackage[latin1]{inputenc}
\RequirePackage[T1]{fontenc}
\RequirePackage{mathrsfs}
\RequirePackage{amsmath}
\RequirePackage{amssymb}
\RequirePackage{amsbsy}
\RequirePackage{bm}
\begin{document}
\title{\bf{On geometric relativistic foundations of\\ matter field equations and\\
plane wave solutions}}
\author{Luca Fabbri\\ 
\footnotesize Dipartimento di Fisica, Universit{\`a} di Bologna,
 Via Irnerio 46, 40126 Bologna, ITALY}
\date{}
\maketitle
\ \ \ \ \ \ \ \ \ \ \ \ \ \ \ \ \ \ \ \ \textbf{PACS}: 04.20.Cv $\cdot$ 04.20.Gz $\cdot$ 04.20.-q
\begin{abstract}
In this paper, we start from the geometric relativistic foundations to define the basis upon which matter field theories are built, and their wave solutions are investigated, finding that they display repulsive interactions able to reproduce the exclusion principle in terms of its effects in a dynamical way, then discussing possible consequences and problems.
\end{abstract}
\section*{Introduction}
In this paper, we shall start from the most general geometric relativistic foundations of the Sciama-Kibble completion of Einsteinian gravitation in presence of the Maxwell electrodynamic field defining the basis upon which the most general Dirac matter field theory will be constructed, and within the matter field equations the torsion-spin coupling field equations result into matter field self-interactions; then we will look for their plane wave solutions, and we will see that they display repulsive interactions: these will be able to entail the exclusion principle by reproducing its effects in a dynamical way. Eventually we study further consequences highlighting possible problems that may arise.
\section{Geometric Relativistic Foundations}
In the geometry of relativistic coordinate tensors, coordinate tensors are defined by their transformation law under coordinate transformations, and as a consequence of the fact that there are two transformations given as direct or inverse then there are two possible different coordinate indices as upper and lower; upper indices are lowered by using the lowering procedure after the introduction of the coordinate tensor $g_{\alpha\sigma}$ as well as lower indices are raised by using the raising procedure after the introduction of the coordinate tensor $g^{\alpha\sigma}$, and these two tensors are symmetric and one the inverse of the other $g_{\nu\rho}g^{\rho\mu} =\delta_{\nu}^{\mu}$ thus they are called coordinate metric tensors. Dynamical properties for coordinate tensors are defined through coordinate covariant derivatives, so that after the introduction of the coordinate connections $\Gamma^{\alpha}_{\mu\nu}$ defined in term of their transformation law, the coordinate covariant derivative $D_{\mu}$ act on coordinate tensors yielding coordinate tensors; the most general coordinate connection is not symmetric in the two lower indices and it has a Cartan torsion tensor $Q_{\alpha\mu\rho}$ which will be considered to be completely antisymmetric $Q_{[\alpha\mu\rho]}=6Q_{\alpha\mu\rho}$ in addition to the metricity condition $D_{\alpha}g=0$ so that the coordinate connection will have one symmetric part written in terms of the coordinate metric tensor, with the consequence that the vanishing of the symmetric part of the coordinate connection and the flattening of the coordinate metric tensor coincide, thus entailing the causality and equivalence principles, as it has been discussed in \cite{f/1,f/2}.

From the coordinate metric tensor the metric completely antisymmetric tensor $\varepsilon$ is defined; with it the completely antisymmetric torsion tensor is the dual of an axial torsion vector defined as $Q_{\mu\rho\beta} =\varepsilon_{\mu\rho\beta\theta}W^{\theta}$ where $D_{\alpha}\varepsilon=0$ identically.

Thus said we have that the coordinate connections $\Gamma^{\alpha}_{\mu\nu}$ can be written as
\begin{eqnarray}
&\Gamma^{\mu}_{\phantom{\mu}\sigma\pi}
=\frac{1}{2}g^{\mu\rho}\left[Q_{\rho\sigma\pi}+\left(\partial_{\pi}g_{\sigma\rho}
+\partial_{\sigma}g_{\pi\rho}-\partial_{\rho}g_{\sigma\pi}\right)\right]
\label{coordinateconnection}
\end{eqnarray}
as decomposed into a coordinate connection $\Lambda^{\alpha}_{\mu\nu}$ symmetric in the two lower indices and verifying the metricity condition plus the torsion, so that the coordinate covariant derivatives $D_{\mu}$ will be separated into the coordinate covariant derivative $\nabla_{\mu}$ plus the additional torsional contribution.

Considering the coordinate connection we define
\begin{eqnarray}
&G^{\mu}_{\phantom{\mu}\rho\sigma\pi}=\partial_{\sigma}\Gamma^{\mu}_{\rho\pi}
-\partial_{\pi}\Gamma^{\mu}_{\rho\sigma}
+\Gamma^{\mu}_{\lambda\sigma}\Gamma^{\lambda}_{\rho\pi}
-\Gamma^{\mu}_{\lambda\pi}\Gamma^{\lambda}_{\rho\sigma}
\label{coordinateRiemann}
\end{eqnarray}
which is a coordinate tensor antisymmetric in the first and second couple of indices called Riemann curvature tensor, and consequently Riemann curvature tensor has one independent contraction $G^{\alpha}_{\phantom{\alpha}\rho\alpha\sigma}=G_{\rho\sigma}$ called Ricci curvature tensor with one contraction $G_{\rho\sigma}g^{\rho\sigma}=G$ called Ricci curvature scalar.

Because of the decomposition above we can write
\begin{eqnarray}
&G^{\mu}_{\phantom{\mu}\rho\sigma\pi}=R^{\mu}_{\phantom{\mu}\rho\sigma\pi}
+\frac{1}{2}(\nabla_{\sigma}Q^{\mu}_{\phantom{\mu}\rho\pi}
-\nabla_{\pi}Q^{\mu}_{\phantom{\mu}\rho\sigma})
+\frac{1}{4}(Q^{\mu}_{\phantom{\mu}\lambda\sigma}Q^{\lambda}_{\phantom{\lambda}\rho\pi}
-Q^{\mu}_{\phantom{\mu}\lambda\pi}Q^{\lambda}_{\phantom{\lambda}\rho\sigma})
\label{coordinateRiemanndecomposition}
\end{eqnarray}
where the torsionless curvature tensor $R^{\alpha}_{\phantom{\alpha}\rho\alpha\sigma}$ is separated from torsion tensor.

With this completely antisymmetric Cartan torsion tensor and Riemann curvature tensor, the commutator of two coordinate covariant derivatives is
\begin{eqnarray}
\nonumber
&[D_{\zeta},D_{\theta}]T^{\alpha...\sigma}
=Q^{\mu}_{\phantom{\mu}\zeta\theta}D_{\mu}T^{\alpha...\sigma}+\\
&+T^{\nu...\sigma}G^{\alpha}_{\phantom{\alpha}\nu\zeta\theta}+...
+T^{\alpha...\nu}G^{\sigma}_{\phantom{\sigma}\nu\zeta\theta}
\end{eqnarray}
showing that the completely antisymmetric Cartan torsion acts as the completely antisymmetric structure coefficients while the Riemann curvature acts as a linear operator, if we interpret the covariant derivatives as generators of a generalized Lie algebra, in which torsion and curvature are the strengths of the translational and rotational potentials of the gauged Poincar\'{e} group \cite{Capozziello:2011et}.

The commutator of three coordinate covariant derivatives in cyclic permutations also gives the expression
\begin{eqnarray}
\nonumber
&(D_{\kappa}Q^{\rho}_{\phantom{\rho}\mu \nu}
+Q^{\rho}_{\phantom{\rho}\kappa \pi}Q^{\pi}_{\phantom{\pi}\mu \nu}
+G^{\rho}_{\phantom{\rho}\kappa\mu\nu})
+(D_{\nu}Q^{\rho}_{\phantom{\rho} \kappa \mu}
+Q^{\rho}_{\phantom{\rho}\nu \pi}Q^{\pi}_{\phantom{\pi}\kappa \mu}
+G^{\rho}_{\phantom{\rho}\nu\kappa\mu})+\\
&+(D_{\mu}Q^{\rho}_{\phantom{\rho} \nu \kappa}
+Q^{\rho}_{\phantom{\rho}\mu \pi}Q^{\pi}_{\phantom{\pi} \nu \kappa}
+G^{\rho}_{\phantom{\rho}\mu\nu\kappa})\equiv0
\end{eqnarray}
called torsion Jacobi-Bianchi identities and
\begin{eqnarray}
\nonumber
&(D_{\mu}G^{\nu}_{\phantom{\nu}\iota \kappa \rho}
-G^{\nu}_{\phantom{\nu}\iota \beta \mu}Q^{\beta}_{\phantom{\beta}\kappa\rho})
+(D_{\kappa}G^{\nu}_{\phantom{\nu}\iota \rho \mu}
-G^{\nu}_{\phantom{\nu}\iota \beta \kappa}Q^{\beta}_{\phantom{\beta}\rho\mu})+\\
&+(D_{\rho}G^{\nu}_{\phantom{\nu}\iota \mu \kappa}
-G^{\nu}_{\phantom{\nu}\iota \beta \rho}Q^{\beta}_{\phantom{\beta}\mu\kappa})\equiv0
\end{eqnarray}
called curvature Jacobi-Bianchi identities, whose contraction is given by
\begin{eqnarray}
&D_{\rho}Q^{\rho\mu\nu}+\left(G^{\nu\mu}-\frac{1}{2}g^{\nu\mu}G\right)
-\left(G^{\mu\nu}-\frac{1}{2}g^{\mu\nu}G\right)\equiv0
\label{torsiondiv}
\end{eqnarray}
called fully contracted torsion Jacobi-Bianchi identities and
\begin{eqnarray}
\nonumber
&D_{\mu}G^{\mu}_{\phantom{\mu}\iota \kappa \rho}
-D_{\kappa}G_{\iota \rho}+D_{\rho}G_{\iota \kappa}+\\
&+G_{\iota \beta}Q^{\beta}_{\phantom{\beta}\kappa\rho}
-G^{\mu}_{\phantom{\mu}\iota \beta \kappa}Q^{\beta}_{\phantom{\beta}\rho\mu}
+G^{\mu}_{\phantom{\mu}\iota \beta \rho}Q^{\beta}_{\phantom{\beta}\kappa\mu}\equiv0
\end{eqnarray}
called contracted curvature Jacobi-Bianchi identities giving
\begin{eqnarray}
&D_{\rho}\left(G^{\rho\kappa}-\frac{1}{2}g^{\rho\kappa}G\right)
+\left(G_{\rho\beta}-\frac{1}{2}g_{\rho\beta}G\right)Q^{\rho\beta\kappa}
+\frac{1}{2}Q_{\nu\rho\beta}G^{\nu\rho\beta\kappa}\equiv0
\label{curvaturediv}
\end{eqnarray}
known as fully contracted curvature Jacobi-Bianchi identities.

Finally we have that
\begin{eqnarray}
&\nabla_{\mu}R^{\nu}_{\phantom{\nu}\iota \kappa \rho}
+\nabla_{\kappa}R^{\nu}_{\phantom{\nu}\iota \rho \mu}
+\nabla_{\rho}R^{\nu}_{\phantom{\nu}\iota \mu \kappa}\equiv0
\end{eqnarray}
are known as the torsionless curvature Jacobi-Bianchi identities.

At this point, it is fundamental to introduce a formally analogous geometry for relativistic complex fields, which are complex fields that are thus defined up to the complex phase $e^{iq\theta}$ where $q$ is the charge label; complex conjugation allows for the inversion of the sign of the charge. The dynamical properties for the complex fields are defined through the gauge covariant derivatives, so that after the introduction of the gauge connection $A_{\mu}$ defined in term of its transformation law, the gauge covariant derivative $D_{\mu}$ acts on complex fields to give complex fields; the gauge covariant derivative will be taken not to contain any torsional contribution, for the reasons that have been explained in \cite{f/1,f/2}.

Now considering the gauge connection it is possible to define
\begin{eqnarray}
&F_{\mu\nu}=\partial_{\mu}A_{\nu}-\partial_{\nu}A_{\mu}
\label{Maxwell}
\end{eqnarray}
being a tensor antisymmetric in the two indices called Maxwell curvature tensor.

With the Maxwell curvature tensor we can write the commutator of two gauge covariant derivatives as 
\begin{eqnarray}
&[D_{\zeta},D_{\theta}]\psi=iqF_{\zeta\theta}\psi
\end{eqnarray}
which is therefore a geometric construction.

We notice in addition that the commutator defined before can be applied to the Maxwell curvature tensor so that after its full contraction we obtain
\begin{eqnarray}
&D_{\rho}\left(D_{\sigma}F^{\sigma\rho}+\frac{1}{2}Q^{\rho\alpha\mu}F_{\alpha\mu}\right)=0
\label{gaugediv}
\end{eqnarray}
as a geometric identity that will be important in what we are going to do next.

Now back to the geometry of relativistic coordinate tensors, we notice that with the coordinate metric tensors $g_{\alpha\sigma}$ and $g^{\alpha\sigma}$ we can define metric concepts like lengths and angles, and so considered a pair of bases of vectors called vierbeins $\xi^{a}_{\sigma}$ and $\xi_{a}^{\sigma}$ dual of one another $\xi^{a}_{\mu}\xi_{a}^{\rho}=\delta^{\rho}_{\mu}$ and $\xi^{a}_{\mu}\xi_{r}^{\mu}=\delta_{r}^{a}$ it is always possible without any loss of generality to write them in such a way that they are orthonormal $\xi^{a}_{\sigma} \xi^{q}_{\rho} g^{\sigma\rho}=\eta^{aq}$ or $\xi_{a}^{\sigma} \xi_{q}^{\rho} g_{\sigma\rho}=\eta_{aq}$ where $\eta_{aq}$ and $\eta^{aq}$ are unitary diagonal matrices called Minkowskian matrices; notice that the vierbeins are determined up to a Lorentz transformation: the introduction of the vierbeins is essential because after multiplying a coordinate tensor by the vierbein and contracting their coordinate indices, we are left with a world index in a world tensor, so that the transformation law for coordinate tensors in terms of the most general coordinates transformation becomes a transformation law for world tensors in terms of a Lorentz transformation. In the geometry of relativistic world tensors, the world tensors are defined in terms of their transformation law under Lorentz transformations, and also in this case the two possible Lorentz transformations give two possible different world indices; the lowering and raising world indices procedure is again possible by means of $\eta_{ab}$ and $\eta^{ab}$, which are symmetric and they are reciprocal of one another $\eta_{ab}\eta^{bm} =\delta_{a}^{m}$ and therefore known as Minkowskian metric matrices. And again the dynamical properties for world tensors are defined through world covariant derivatives, so that after the introduction of the Lorentz connection given by $\Gamma^{i}_{j\mu}$ and defined by its transformation law, the world covariant derivative $D_{\mu}$ acts on world tensors giving world tensors; notice that an analogous of the torsion tensor cannot be defined for the connection in a formalism in which the two lower indices are of different types, while instead the condition $D_{\alpha}\xi=0$ is imposed as before to make the coordinate and world covariant derivatives coincide and the vanishing of the covariant derivative of the Minkowskian metric matrices $D_{\alpha}\eta=0$ is taken to represent the fact that these matrices are constant in the present formalism.

From the coordinate metric tensor we can define the metric completely antisymmetric tensor $\varepsilon$ as usual, but which will now be given in terms of the vierbein; on the other hand the relationship $D_{\alpha}\varepsilon=0$ still holds but now it is trivial because the tensor $\varepsilon$ is constant in the present formalism as well.

The two metricity conditions together imply that the Lorentz connection is to be written as follows
\begin{eqnarray}
&\Gamma^{b}_{\phantom{b}j\mu}=
\xi^{\alpha}_{j}\xi_{\rho}^{b}\left(\Gamma^{\rho}_{\phantom{\rho}\alpha\mu}
+\xi_{\alpha}^{k}\partial_{\mu}\xi^{\rho}_{k}\right)
\label{Lorentzconnection}
\end{eqnarray}
in terms of the coordinate connection and it is antisymmetric in its world indices.

Considering the Lorentz connection it is possible to define
\begin{eqnarray}
&G^{a}_{\phantom{a}b\sigma\pi}
=\partial_{\sigma}\Gamma^{a}_{b\pi}-\partial_{\pi}\Gamma^{a}_{b\sigma}
+\Gamma^{a}_{j\sigma}\Gamma^{j}_{b\pi}-\Gamma^{a}_{j\pi}\Gamma^{j}_{b\sigma}
\label{LorentzRiemann}
\end{eqnarray}
which is a tensor antisymmetric in both the coordinate and the world indices.

We have that this tensor is writable as
\begin{eqnarray}
&G^{a}_{\phantom{a}b\sigma\pi}=G^{\mu}_{\phantom{\mu}\rho\sigma\pi}\xi^{\rho}_{b}\xi^{a}_{\mu}
\label{correlation}
\end{eqnarray}
in terms of the previous expression of the Riemann curvature tensor.

With this expression the commutator of two covariant derivatives is
\begin{eqnarray}
\nonumber
&[D_{\zeta},D_{\theta}]T^{\alpha...\sigma a...s}=
Q^{\mu}_{\phantom{\mu}\zeta\theta}D_{\mu}T^{\alpha...\sigma a...s}+\\
&+T^{\alpha...\sigma j...s}G^{a}_{\phantom{a}j\zeta\theta}+...
+T^{\alpha...\sigma a...j}G^{s}_{\phantom{s}j\zeta\theta}
\end{eqnarray}
still having the covariant derivatives as generators of a generalized Lie algebra.

Now recall that the introduction of the geometry of relativistic complex fields was done in view of possible applications to complex fields; and the translation to the vierbein formalism is important because Lorentz transformations have a structure that can be made explicit, and consequently its structure can also be given as written in terms of other representations: of all the possible Lorentz group representations then, the most special are the complex representations, built by considering a set of complex matrices $\boldsymbol{\gamma}_{a}$ belonging to the Clifford complex algebra $\{\boldsymbol{\gamma}_{i},\boldsymbol{\gamma}_{j}\}=2\boldsymbol{\mathbb{I}}\eta_{ij}$ verifying the following relationships
\begin{eqnarray}
&\boldsymbol{\gamma}_{i}\boldsymbol{\gamma}_{j}\boldsymbol{\gamma}_{k}
=\boldsymbol{\gamma}_{i}\eta_{jk}-\boldsymbol{\gamma}_{j}\eta_{ik}
+\boldsymbol{\gamma}_{k}\eta_{ij}+i\varepsilon_{ijkq}\boldsymbol{\gamma}\boldsymbol{\gamma}^{q}
\end{eqnarray}
so that we can build $[\boldsymbol{\gamma}_{i},\boldsymbol{\gamma}_{j}]=4\boldsymbol{\sigma}_{ij}$ and define $\{\boldsymbol{\gamma}_{i},\boldsymbol{\sigma}_{jk}\}= i\varepsilon_{ijkq}\boldsymbol{\gamma}\boldsymbol{\gamma}^{q}$ verifying
\begin{eqnarray}
&[\boldsymbol{\gamma}_{i},\boldsymbol{\sigma}_{jk}]
=\eta_{ij}\boldsymbol{\gamma}_{k}-\eta_{ik}\boldsymbol{\gamma}_{j}\\
&[\boldsymbol{\sigma}_{ab},\boldsymbol{\sigma}_{jk}]
=\eta_{ak}\boldsymbol{\sigma}_{bj}-\eta_{aj}\boldsymbol{\sigma}_{bk}
-\eta_{bk}\boldsymbol{\sigma}_{aj}+\eta_{bj}\boldsymbol{\sigma}_{ak}
\end{eqnarray}
as the complex generators of the infinitesimal form of the complex Lorentz transformation $\boldsymbol{S}$ called spinorial transformation. Once the explicit form of the spinorial transformation $\boldsymbol{S}$ is given, we can define the complex fields that transform according to this transformation as spinor fields, classified in terms of half-integer spin: in what follows we will restrict ourselves to the case given by the simplest $\frac{1}{2}$-spin spinor field. In this geometry of relativistic spinor fields, the spinor fields are defined as transforming according to the spinorial transformation law with $\frac{1}{2}$-spin label; as the transformation $\boldsymbol{S}$ may be either direct or inverse then there are two types of spinors, related by the adjoint procedure given in terms of the matrix $\boldsymbol{\gamma}_{0}$ as $\overline{\psi}\equiv\psi^{\dagger}\boldsymbol{\gamma}_{0}$ and vice-versa. The dynamical properties for spinor fields are defined through covariant derivatives, so that after the introduction of the spinorial connection $\boldsymbol{\Gamma}_{\mu}$ defined in terms of its transformation law, the spinorial covariant derivative $\boldsymbol{D}_{\mu}$ acts on spinor fields giving spinor fields; then the $\boldsymbol{\gamma}_{j}$ and $\boldsymbol{\sigma}_{ij}$ matrices are constant by construction.

Now considering the fact that the $\boldsymbol{\gamma}_{j}$ and $\boldsymbol{\sigma}_{ij}$ matrices are constant it is possible to see that the most general spinorial connection is given by
\begin{eqnarray}
&\boldsymbol{A}_{\mu}=\frac{1}{2}\Gamma^{ab}_{\phantom{ab}\mu}\boldsymbol{\sigma}_{ab}+
iqA_{\mu}\boldsymbol{\mathbb{I}}
\label{spinorialconnection}
\end{eqnarray}
in terms of the Lorentz connection plus an abelian field which we may now identify with the gauge connection we have introduced previously.

Considering this spinorial connection it is possible to define
\begin{eqnarray}
&\boldsymbol{F}_{\sigma\pi}
=\partial_{\sigma}\boldsymbol{A}_{\pi}-\partial_{\pi}\boldsymbol{A}_{\sigma}
+[\boldsymbol{A}_{\sigma},\boldsymbol{A}_{\pi}]
\label{RiemannMaxwell}
\end{eqnarray}
which is a tensorial spinor antisymmetric in the tensorial indices.

This tensorial spinor is writable as
\begin{eqnarray}
&\boldsymbol{F}_{\sigma\pi}=\frac{1}{2}G^{ab}_{\phantom{ab}\sigma\pi}\boldsymbol{\sigma}_{ab}
+iqF_{\sigma\pi}\boldsymbol{\mathbb{I}}
\label{RiemannMaxwellcombination}
\end{eqnarray}
as a combination of both the Riemann and Maxwell curvature tensors.

With this compact expression then the commutator of two gauge covariant derivatives is given by the following
\begin{eqnarray}
&[\boldsymbol{D}_{\zeta},\boldsymbol{D}_{\theta}]\psi
=Q^{\mu}_{\phantom{\mu}\zeta\theta}\boldsymbol{D}_{\mu}\psi+\boldsymbol{F}_{\zeta\theta}\psi
\label{covariantgaugecommutator}
\end{eqnarray}
which is a geometric construction important in what we are going to do next.

We notice that the only hypotheses assumed were arguments of symmetry for spacetime-gauge transformations in these geometric relativistic foundations.

Now we have that the Jacobi-Bianchi identities in their fully contracted form for the torsion (\ref{torsiondiv}) and for the curvature (\ref{curvaturediv}) and also the identity of the commutator of the Maxwell tensor in its fully contracted form (\ref{gaugediv}) are satisfied through the identity of the commutator of the spinor field (\ref{covariantgaugecommutator}) if we postulate the system of field equations given by the torsion coupling to the spin
\begin{eqnarray}
&Q^{\rho\mu\nu}
=-\frac{i}{4}\overline{\psi}\{\boldsymbol{\gamma}^{\rho},\boldsymbol{\sigma}^{\mu\nu}\}\psi
\end{eqnarray}
and the curvature coupling to the energy
\begin{eqnarray}
&G^{\mu}_{\phantom{\mu}\nu}-\frac{1}{2}\delta^{\mu}_{\nu}G
-\frac{1}{8}\delta^{\mu}_{\nu}F^{2}+\frac{1}{2}F^{\rho\mu}F_{\rho\nu}
=\lambda\delta^{\mu}_{\nu}+
\frac{i}{4}\left(\overline{\psi}\boldsymbol{\gamma}^{\mu}\boldsymbol{D}_{\nu}\psi
-\boldsymbol{D}_{\nu}\overline{\psi}\boldsymbol{\gamma}^{\mu}\psi\right)
\end{eqnarray}
and also the gauge field coupling to the current
\begin{eqnarray}
&D_{\sigma}F^{\sigma\rho}+\frac{1}{2}F_{\mu\nu}Q^{\mu\nu\rho}
=q\left(\overline{\psi}\boldsymbol{\gamma}^{\rho}\psi\right)
\end{eqnarray}
together with the matter field equation
\begin{eqnarray}
&i\boldsymbol{\gamma}^{\mu}\boldsymbol{D}_{\mu}\psi
-\left(m\boldsymbol{\mathbb{I}}+ib\boldsymbol{\gamma}\right)\psi=0
\end{eqnarray}
in terms of the system of integration constants $\lambda$, $m$ and $b$, as discussed in \cite{f/3,f/4}.

In this system of field equations it is possible to separate torsion and because the field equation for the torsion coupling to the spin is algebraic then we can substitute torsion with the spin written in terms of the matter field to get
\begin{eqnarray}
\nonumber
&R_{\mu\nu}-\frac{1}{8}g_{\mu\nu}F^{2}+\frac{1}{2}g^{\eta\rho}F_{\mu\eta}F_{\nu\rho}
=-\lambda g_{\mu\nu}-\frac{1}{4}\left[m\left(\overline{\psi}\psi\right)
+b\left(i\overline{\psi}\boldsymbol{\gamma}\psi\right)\right]g_{\mu\nu}+\\
&+\frac{i}{8}\left(\overline{\psi}\boldsymbol{\gamma}_{\mu}\boldsymbol{\nabla}_{\nu}\psi
+\overline{\psi}\boldsymbol{\gamma}_{\nu}\boldsymbol{\nabla}_{\mu}\psi
-\boldsymbol{\nabla}_{\nu}\overline{\psi}\boldsymbol{\gamma}_{\mu}\psi
-\boldsymbol{\nabla}_{\mu}\overline{\psi}\boldsymbol{\gamma}_{\nu}\psi\right)
\end{eqnarray}
and
\begin{eqnarray}
&\nabla_{\sigma}F^{\sigma\rho}=q\left(\overline{\psi}\boldsymbol{\gamma}^{\rho}\psi\right)
\end{eqnarray}
along with
\begin{eqnarray}
\nonumber
&i\boldsymbol{\gamma}^{\mu}\boldsymbol{\nabla}_{\mu}\psi
-\frac{3}{16}\left[\left(\overline{\psi}\psi\right)\boldsymbol{\mathbb{I}}
+i\left(i\overline{\psi}\boldsymbol{\gamma}\psi\right)\boldsymbol{\gamma}\right]\psi
-\left(m\boldsymbol{\mathbb{I}}+ib\boldsymbol{\gamma}\right)\psi\equiv\\
&\equiv i\boldsymbol{\gamma}^{\mu}\boldsymbol{\nabla}_{\mu}\psi
-\frac{3}{16}\left(\overline{\psi}\boldsymbol{\gamma}_{\mu}\psi\right)\boldsymbol{\gamma}^{\mu}\psi
-\left(m\boldsymbol{\mathbb{I}}+ib\boldsymbol{\gamma}\right)\psi=0
\end{eqnarray}
where the field equations for the curvature-energy coupling are equal to the field equations for the curvature-energy coupling we would have had in the torsionless case and the field equations for the gauge-current coupling are equal to the field equations for the gauge-current coupling we have in the torsionless case but the field equations for the matter field are equivalent to the equations for the matter field in the torsionless case plus torsional potentials, which are given by specific fields that are constructed in terms of the matter fields themselves.

We remark that starting from the previous geometric relativistic foundations, the only hypotheses assumed were requirements of having the background coupled to its matter content in terms of background-matter field equations, in order to consistently build these matter field theories.
\subsection{Matter Field Equations}
Because of the causality and equivalence principle, there exists one system of coordinates in which locally the metric is constant and the connection vanishes so that the whole connection is negligible and therefore the covariant derivative is the partial derivative calculated with respect to the spacetime position that is expressed as in the ordinary circumstance of macroscopic situations.

First of all we notice that for the macroscopic situation we want to study, the field equations for the curvature-energy coupling get the form
\begin{eqnarray}
\nonumber
&R_{\mu\nu}-\frac{1}{8}g_{\mu\nu}F^{2}+\frac{1}{2}g^{\eta\rho}F_{\mu\eta}F_{\nu\rho}
=-\left[\lambda-\frac{3}{64}\left(\overline{\psi}\psi\right)^{2}\right]g_{\mu\nu}-\\
&-\frac{1}{4}\left[m+\frac{3}{16}\left(\overline{\psi}\psi\right)\right]
\left(\overline{\psi}\psi\right)g_{\mu\nu}
+\frac{1}{2}\left[m+\frac{3}{16}\left(\overline{\psi}\psi\right)\right]
\left(\overline{\psi}\psi\right)U_{\mu}U_{\nu}
\label{gravitational}
\end{eqnarray}
where the additional potentials may be written as shifts for the integration constants given by $64\Lambda=64\lambda-3(\overline{\psi}\psi)^{2}$ and $16M=16m+3(\overline{\psi}\psi)$ so that once the mass density is defined to be given by $\mu=M(\overline{\psi}\psi)$ we obtain the field equations for the curvature-energy coupling as in the macroscopic situation.

For the field equations of the gauge-current coupling we have
\begin{eqnarray}
&\nabla_{\sigma}F^{\sigma\rho}=q\left(\overline{\psi}\psi\right)U^{\rho}
\label{electrodynamic}
\end{eqnarray}
where once the charge density is $\rho=q(\overline{\psi}\psi)$ we obtain the field equations for the gauge-current coupling recovering the limit for the macroscopic situation.

Finally for the matter field equations we have that they are
\begin{eqnarray}
\nonumber
&i\boldsymbol{\gamma}^{\mu}\boldsymbol{\nabla}_{\mu}\psi
-\frac{3}{16}\left(\overline{\psi}\psi\right)\psi-m\psi\equiv\\
&\equiv i\boldsymbol{\gamma}^{\mu}\boldsymbol{\nabla}_{\mu}\psi
-\frac{3}{16}\left(\overline{\psi}\boldsymbol{\gamma}_{\mu}\psi\right)
\boldsymbol{\gamma}^{\mu}\psi-m\psi=0
\label{matter}
\end{eqnarray}
where the non-linear term has the scalar form of a mass term or equivalently the vectorial form of an electrodynamic potential, so that the dynamics is qualitatively that of a matter field with the mass term in the dispersion relations or equivalently interacting with its own electrodynamic field with a diffusing back-reaction, and since these ones are the Dirac matter field equations with the Nambu-Jona--Lasinio or Gross-Neveu or Thirring repulsive self-interactions, then the spreading of the matter field all over the whole spacetime is to be expected, as it is discussed by these authors respectively in \cite{n-j--l/1,n-j--l/2}, \cite{g-n} and \cite{t}.

When in matter field equations (\ref{matter}) we separate the spatial and the temporal components it becomes possible to take the stationary configuration in the limit of small velocities and in the limit of small values of the spinorial field, so that the small semispinorial component tends to vanish with the large semispinorial component verifying the approximated matter field equation
\begin{eqnarray}
\nonumber
&i\frac{\partial\phi}{\partial t}
+\frac{1}{2m}\boldsymbol{\sigma}^{k}\boldsymbol{\nabla}_{k}
\boldsymbol{\sigma}^{a}\boldsymbol{\nabla}_{a}\phi
-\frac{3}{16}\left(\phi^{\dagger}\phi\right)\phi\equiv\\
&\equiv i\frac{\partial\phi}{\partial t}
+\frac{1}{2m}\boldsymbol{\sigma}^{k}\boldsymbol{\nabla}_{k}
\boldsymbol{\sigma}^{a}\boldsymbol{\nabla}_{a}\phi
-\frac{3}{16}\left(\phi^{\dagger}\boldsymbol{\sigma}^{a}\phi\right)
\boldsymbol{\sigma}_{a}\phi=0
\label{matterapproximated} 
\end{eqnarray}
where the non-linear term has the form of a rotational potential, so that the dynamics is qualitatively that of a matter field scattered around by the centrifugal barrier of its self-interaction, and as these are the Schr\"{o}dinger matter field equations with Ginzburg-Landau or Heisenberg repulsive self-interactions, the spread of the matter field over space is expected, as discussed in \cite{g-l} and \cite{h}.

To summarize, we have that, within the Dirac or the correspondingly approximated Schr\"{o}dinger field equations, the torsion tensor can be separated, then converted, through the torsion-spin coupling field equations, into spinorial self-interactions which, with suitable rearrangements, are rewritten in the form Nambu-Jona--Lasinio or Gross-Neveu or Thirring or the correspondingly approximated Ginzburg-Landau or Heisenberg repulsive self-interactions, and thus the spreading of the matter field has to be expected.

Now in order for the curvature-energy and gauge-current coupling field equations to be written in the form (\ref{gravitational}) and (\ref{electrodynamic}) the matter field equations written in the form (\ref{matter}) must have plane wave solutions in chiral representation as
\begin{eqnarray}
&\psi=\mathrm{e}^{-ix^{\mu}P_{\mu}\left(1+\frac{3A^{2}}{16m}\right)}
\left[\begin{tabular}{c}
$\sqrt{\frac{E-P}{2}}\sqrt{\frac{A^{2}}{m}}\cos{\frac{\theta}{2}}$\\
$\sqrt{\frac{E+P}{2}}\sqrt{\frac{A^{2}}{m}}\sin{\frac{\theta}{2}}$\\
$\sqrt{\frac{E+P}{2}}\sqrt{\frac{A^{2}}{m}}\cos{\frac{\theta}{2}}$\\
$\sqrt{\frac{E-P}{2}}\sqrt{\frac{A^{2}}{m}}\sin{\frac{\theta}{2}}$
\end{tabular}\right]
\label{solutionchiral}
\end{eqnarray}
which is suitable to study the high-energy approximation given by
\begin{eqnarray}
&\psi=\mathrm{e}^{-ix^{\mu}P_{\mu}\left(1+\frac{3A^{2}}{16m}\right)}
\left[\begin{tabular}{c}
$0$\\
$\sqrt{E}\sqrt{\frac{A^{2}}{m}}\sin{\frac{\theta}{2}}$\\
$\sqrt{E}\sqrt{\frac{A^{2}}{m}}\cos{\frac{\theta}{2}}$\\
$0$
\end{tabular}\right]
\label{solutionapproximatedhigh}
\end{eqnarray}
in which the helicity eigenstates are well defined and in which it is therefore possible to see that this solution is the superposition of two solutions that are eigenstates with opposite eigenvalues of the helicity operator, or alternatively it is also possible to write the solution in the standard representation as
\begin{eqnarray}
&\psi=\mathrm{e}^{-ix^{\mu}P_{\mu}\left(1+\frac{3A^{2}}{16m}\right)}
\left[\begin{tabular}{c}
$\left(\sqrt{\frac{E+P}{4}}+\sqrt{\frac{E-P}{4}}\right)
\sqrt{\frac{A^{2}}{m}}\cos{\frac{\theta}{2}}$\\
$\left(\sqrt{\frac{E-P}{4}}+\sqrt{\frac{E+P}{4}}\right)
\sqrt{\frac{A^{2}}{m}}\sin{\frac{\theta}{2}}$\\
$\left(\sqrt{\frac{E+P}{4}}-\sqrt{\frac{E-P}{4}}\right)
\sqrt{\frac{A^{2}}{m}}\cos{\frac{\theta}{2}}$\\
$\left(\sqrt{\frac{E-P}{4}}-\sqrt{\frac{E+P}{4}}\right)
\sqrt{\frac{A^{2}}{m}}\sin{\frac{\theta}{2}}$
\end{tabular}\right]
\label{solutionstandard}
\end{eqnarray}
which is suitable to get the low-energy approximation given by
\begin{eqnarray}
&\psi=\mathrm{e}^{-ix^{\mu}P_{\mu}\left(1+\frac{3A^{2}}{16m}\right)}
\left[\begin{tabular}{c}
$\sqrt{E}\sqrt{\frac{A^{2}}{m}}\cos{\frac{\theta}{2}}$\\
$\sqrt{E}\sqrt{\frac{A^{2}}{m}}\sin{\frac{\theta}{2}}$\\
$0$\\
$0$
\end{tabular}\right]
\label{solutionapproximatedlow}
\end{eqnarray}
still being the superposition of two solutions that are eigenstates with opposite eigenvalues of the projection along the third component of the spin operator, and where $x^{\mu}$ is the spacetime position with $A$ a scale factor and $\theta$ the angle between the spin and the momentum that is taken to be along the third axis.

The approximated matter field equations have plane wave solutions that can also be obtained from the approximation of the plane wave solutions as
\begin{eqnarray}
&\phi=\mathrm{e}^{-i\left[t\left(\frac{P^{2}}{2m}+\frac{3A^{2}}{16}\right)-x^{a}P_{a}\right]}
\left[\begin{tabular}{c}
$A\cos{\frac{\theta}{2}}$\\
$A\sin{\frac{\theta}{2}}$
\end{tabular}\right]
\label{solutionapproximatedslow}
\end{eqnarray}
which is the superposition of two solutions being eigenstates with opposite eigenvalues of the projection along the third component of the spin operator, and where $t$ is the time and $x^{a}$ the space position with $A$ a scale factor and $\theta$ the angle between the spin and the momentum that is taken along the third axis.

We may summarize what we did by saying that, compatibly with the fact that for repulsive dynamics the separation between two matter fields was expected, the analysis of the plane wave solutions eventually in their approximated form revealed that, because of the $A^{2}$ correction, there is no linear superposition between two wave solutions, although the fact that one such solution may be seen as a linear superposition of two solutions of antialigned spin also indicates that there is linear superposition of two solutions with antialigned spin.
\subsubsection{Plane Wave Solutions}
Having arrived to this point, let us look back to what we have done reconsidering that the matter field equations above had wave solutions in the form
\begin{eqnarray}
&\psi=\mathrm{e}^{-ix^{\mu}P_{\mu}\left(1+\frac{3A^{2}}{16m}\right)}
\left[\begin{tabular}{c}
$\left(\sqrt{\frac{E+P}{4}}+\sqrt{\frac{E-P}{4}}\right)
\sqrt{\frac{A^{2}}{m}}\cos{\frac{\theta}{2}}$\\
$\left(\sqrt{\frac{E-P}{4}}+\sqrt{\frac{E+P}{4}}\right)
\sqrt{\frac{A^{2}}{m}}\sin{\frac{\theta}{2}}$\\
$\left(\sqrt{\frac{E+P}{4}}-\sqrt{\frac{E-P}{4}}\right)
\sqrt{\frac{A^{2}}{m}}\cos{\frac{\theta}{2}}$\\
$\left(\sqrt{\frac{E-P}{4}}-\sqrt{\frac{E+P}{4}}\right)
\sqrt{\frac{A^{2}}{m}}\sin{\frac{\theta}{2}}$
\end{tabular}\right]
\label{solutionnormal}
\end{eqnarray}
or in the limit above the approximated wave solution in the form
\begin{eqnarray}
&\phi=\mathrm{e}^{-i\left[t\left(\frac{P^{2}}{2m}+\frac{3A^{2}}{16}\right)-x^{a}P_{a}\right]}
\left[\begin{tabular}{c}
$A\cos{\frac{\theta}{2}}$\\
$A\sin{\frac{\theta}{2}}$
\end{tabular}\right]
\label{solutionlimit}
\end{eqnarray}
recollecting all the results we have obtained; if we read the meaning of the wave solutions as given by (\ref{solutionnormal}) or (\ref{solutionlimit}) we see that the non-linear potential producing the repulsion that keeps apart two solutions in general fails to do so by allowing the linear superposition of two solutions in the special case for which the two solutions have opposite spins. Thus the presence of torsion with its torsion-spin coupling endows the spinor fields with repulsive self-interactions, hence entailing the exclusion principle by giving rise to its effects dynamically, as it has also been discussed by Sivaram and Sinha and by Sachs in references \cite{s-s} and \cite{s}.

Notice that under the discrete transformation combining spacetime inversion and charge-conjugation
\begin{eqnarray}
&x^{\mu}\rightarrow-x^{\mu}\\
\nonumber
&\psi\rightarrow\boldsymbol{\gamma}\psi
\label{CPtransformation}
\end{eqnarray}
the solutions given by expressions (\ref{solutionnormal}) are not transformed into other solutions because of the presence of the $A^{2}$ correction for the $\psi$ spinor, although after taking the limit above the discrete transformation can only be recovered for the purely spatial inversion
\begin{eqnarray}
&x^{a}\rightarrow-x^{a}
\label{Ptransformation}
\end{eqnarray}
with no transformation for the spinor since after its splitting into semispinorial components the small one has vanished and only the large one is still present in the form of the approximated solution given by expression (\ref{solutionlimit}) now transforming into itself despite the fact that the $A^{2}$ correction is still present. This suggests that in general there may be no matter/antimatter duality albeit in this limit the approximated duality is recovered, as also discussed by Pop{\l}awski in \cite{p}.

Moreover in this limit it is possible to calculate the energy of the field by computing the temporal derivative of the solution to get
\begin{eqnarray}
&E=m+\frac{3A^{2}}{16}
\label{mass}
\end{eqnarray}
or the slow-speed weak-torsion approximation
\begin{eqnarray}
&E=\frac{P^{2}}{2m}+\frac{3A^{2}}{16}
\label{energy}
\end{eqnarray}
which is the sum of the kinetic energy of a particle of mass $m$ plus the kinetic energy of a wave with amplitude $A$, the former term being relevant for small contributions while the latter term being relevant for large contributions of the self-interaction: intriguingly this complementarity between particles and waves is the result of the balance between their two contributions to the kinetic energy, so that what appears to be a particle at low kinetic energies turns its appearance into that of a wave for high kinetic energies; we have also to notice in addition that in this form the kinetic energy is clearly positive, and so well defined.

We notice that it is an already known result that matter fields with mass terms tend to spread throughout the spacetime, and the intrinsically repulsive self-interactions that are due to the interaction with their electrodynamic potentials and even more to the presence of torsion do contribute to this spreading, if we consider their effects within the matter field equation alone; on the other hand however, it is also known that mass, electrodynamics and torsion all contribute to the field energy density that is the source of the gravitational attraction exerted by the field onto itself: the balance between the field self-repulsion and its own gravity may provide the conditions for the existence of stable localized solutions. Then we have that for these stable localized solutions, fluctuations may occur around the minimum of the potential in this compact region, and this circumstance is known to result in general into the discretization of the field energy spectrum; this situation seems to constitute some sort of generalization of the de Broglie quantization already employed to explain the stability of atoms. And this may also imply that a single fundamental matter field with many different mass excitations can look like many resonant matter fields, the one with the lowest mass identified with the stable particle and all those with higher masses identified with all unstable particles of the same family.

To conclude we would like to highlight that by releasing the normalization of the Newton constant $8\pi G=1$ it becomes clear that all effects coming from torsion are relevant only close to the Planck scale, unless the torsional constant is allowed to get a value different from that of the gravitational constant, which may be done either by assuming that there merely are two distinct constants, as suggested by Kaempffer \cite{k}, or by interpolating two values of the same constant with a running coupling, as suggested by Dirac \cite{d}, but we still do not know whether any of these two possible mechanisms may be viable.
\section*{Conclusion}
In this paper, we have been employing only geometric relativistic fundamental identities to get the system of field equations of torsional-metric gravity with electrodynamics for the matter field equations, first in the form of Dirac matter field equations and then the limit given by the Schr\"{o}dinger matter field equations, having Nambu-Jona--Lasinio or Gross-Neveu or Thirring repulsive self-interaction and the limit given by Ginzburg-Landau or Heisenberg repulsive self-interaction; then we have found their plane wave solutions, which have been shown to display corresponding repulsive behaviour: this repulsive behaviour has been seen to entail the exclusion principle by reproducing its effects in a dynamical way, and we have discussed consequences for the lack of duality between matter/antimatter and for the energy of the matter field. Eventually we have sketched a discussion regarding the problem of stability, discretization of the spectrum and classification of particles, finally addressing the issue of the energy scale, indicating possible ways out, about which some are already being pursued in a following work.

\end{document}